\documentclass[]{interact}

\usepackage{epstopdf}% To incorporate .eps illustrations using PDFLaTeX, etc.
\usepackage{subfigure}% Support for small, `sub' figures and tables
\usepackage{xcolor}
\usepackage[numbers,sort&compress,merge]{natbib}% Citation support using natbib.sty
\bibpunct[, ]{(}{)}{,}{n}{,}{,}% Citation support using natbib.sty
\theoremstyle{plain}% Theorem-like structures

\theoremstyle{definition}

\theoremstyle{remark}

\begin{document}

\articletype{}

\title{Mimicking a hybrid-optomechanical system using an intrinsic quadratic coupling in conventional optomechanical system}

\author{
\name{U.~Satya Sainadh\textsuperscript{a} and M.~Anil Kumar\thanks{CONTACT M.~Anil Kumar Email: anilksp.m@gmail.com}\textsuperscript{b}}
\affil{\textsuperscript{a}Center for Quantum Dynamics, Griffith University, Brisbane, QLD 4111, Australia. \textsuperscript{b}National Institute of Technology, Andhra Pradesh, Tadepalligudam, 534101, India.}
}

\maketitle

\begin{abstract}
We consider an optical and mechanical mode interacting through  both linear and quadratic dispersive couplings in a general cavity-optomechanical set-up. The parity and strength of an intrinsic quadratic optomechanical coupling (QOC) provides an opportunity to control the optomechanical (OM) interaction. We quantify this interaction by studying  normal-mode splitting (NMS) as a function of the QOC's strength. The proposed scheme exhibits NMS features equivalent to a hybrid-OM system containing either an optical parametric amplifier (OPA) or a Kerr medium. Such a system in reality could offer an alternative platform for devising state-of-art quantum devices with requiring no extra degrees-of-freedom as in hybrid-OM systems. 
\end{abstract}


\section{Introduction}

Governing mechanical degrees of freedom of a macroscopic object with electromagnetic radiation field (light) using an optomechanical (OM) set-up is a wide popular study. In general for an OM system as shown in Fig. 1, the interaction mediated by radiation pressure force between an optical and mechanical mode is considered as a dispersive OM coupling and can be understood as a single photon-phonon interaction. Here, the interaction is due to the linear displacement ($x$ in its first order) of the mechanical oscillator couples to the cavity frequency $\omega_c$ and is called as the linear optomechanical coupling (LOC). The LOC interaction between optical and mechanical modes opens the possibility of studying and observing some  fundamental quantum effects in mesoscopic systems \cite{aspelmeyer,meystre} that are of great interest from conceptual and as well as fundamental point of view in quantum physics.

It is crucial to either increase the strength of the OM coupling or cool the mechanical oscillator in order to observe quantum features. However, it is experimentally challenging to increase the strength of LOC in an OM system and therefore most of the studies were done by cooling the mechanical oscillator in a resolved side band limit ($\omega_m>\kappa$). In this limit, ground state cooling of the mechanical oscillator \cite{arcizet} would yield normal mode splitting (NMS) \cite{dobrindt,groblacher2009,rossi}. It is similar to the vacuum Rabi splitting observed in cavity transmission that appears when the atoms and cavity field are strongly coupled to each other \cite{eberly,thompson,agarwal}. This feature can be observed by examining the mechanical oscillator's position spectrum and the splitting of the spectrum into a doublet can be used as an indicator for the existence of a strong photon-phonon interaction between optical and mechanical mode. 

In addition to LOC, novel design and development of hybrid-OM systems were widely investigated to explore quantum features and its applications \cite{hybrid,genes2008,genes2011,wang2014,chen}. In these systems, few extra degrees-of-freedom were introduced in the cavity to interact with both optical and mechanical modes, that could create or manipulate the strong OM interaction. Such schemes were proposed with an OM system containing optical parametric amplifier (OPA) enhancing NMS \cite{huangnms}. Whereas on the other hand inserting a strong third order nonlinear Kerr medium ($\chi^3$) inside an OM system, inhibits the NMS \cite{kumar} through photon-blockade mechanism and offer a coherently controlled dynamics for mechanical mode. Later, a hybrid-OM scheme with both OPA and Kerr medium was proposed to control OM interactions \cite{naderi} resulting in a similar conclusion as in \cite{huangnms,kumar}. 

The present report proposes an alternative scheme that uses dispersive couplings both in its linear and quadratic form. Analogous to LOC, the quadratic optomechanical coupling (QOC) is a second order photon-phonon interaction manifested in terms of $x^2$ of the mechanical oscillator as observed in various OM systems \cite{thompson2008,sankey,purdy, kaviani,brawley,paraso}. Such systems can offer mechanical oscillator's cooling and squeezing \cite{girvin}, generation of non-classical states \cite{tan2013}, photon blockade mechanism \cite{liao2013} and features pertaining to the cross correlations between photons and phonons \cite{xu2018}. The present scheme consisting of both LOC and QOC in a OM system as proposed by authors, previously exhibited quantum features like squeezed states of optical \cite{anilkumar} and mechanical modes \cite{satya}. Here in the current report we consider a theoretical study of NMS in a conventional OM system in conjunction with a QOC, with which we show that such a system resembles a hybrid-OM system \cite{huangnms,kumar,naderi}.

The paper is organized as follows: Section II contains a description of system Hamiltonaian, quantum Langevin equations for the systems operators and estimation of steady state mean values. In Section III, we present quantum fluctuations and analysis of the NMS. Section IV discusses the numerical results and finally the conclusions are presented in Section V. 

\section{Theoretical model}
\begin{figure}[t] \centering
 \includegraphics[scale=0.45]{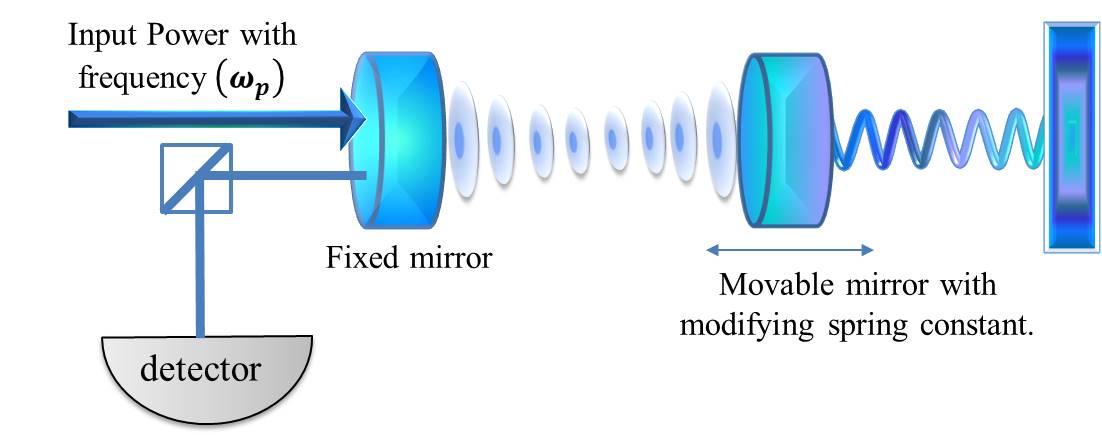}
 \caption{\label{fig1}   Schematic of a cavity-optomechanical system with an optical mode supported in the cavity of line width $\kappa$ and resonance frequency $\omega_c$, driven by an external laser of frequency $\omega_p$ and power $\mathcal{P}$. The radiation pressure interacting through linear and quadratic dispersive couplings exert force on the movable mirror (treated as a mechanical oscillator) that softens or hardens the spring, modifying its natural frequency $\omega_m$ and intrinsic damping rate $\gamma_m$.}
 \end{figure}

 We consider an OM system as illustrated in Fig. \ref{fig1}, with a single cavity mode frequency $\omega_c$ and cavity decay rate $\kappa$ coupled to a single mechanical mode of frequency $\omega_m$, driven by a strong pump field of frequency $\omega_p$. The interactions are both linear $(g_{_{l}})$ and quadratic $(g_{_{q}})$ in mechanical oscillator's displacement. The Hamiltonian of the system expressed using the dimensionless operators of position (\textbf{x}), momentum (\textbf{p}) and cavity annihilation  operator (\textbf{a}) in the laser frame is 
 \small
 \begin{equation}
 \mathbf{H}=\hbar\Delta \mathbf{a^\dagger a}+\frac{\hbar \omega_m }{2}(\mathbf{x}^2+\mathbf{p}^2)+\hbar g_{_{l}} \mathbf{a^\dagger a x}+\hbar g_{_{q}} \mathbf{a^\dagger a}\mathbf{x}^2+i\hbar\mathbf{\varepsilon(a^\dagger-a)} \label{a1}
 \end{equation}\normalsize
such that $[\mathbf{x},\mathbf{p}]=i$, $[\mathbf{a,a^\dagger} ] = 1$ and  $\Delta=\omega_c-\omega_p$ is the cavity detuning. While the first two terms express the free energy of the optical field and mechanical oscillator, the next two terms describe the LOC and QOC interactions. 
The LOC and QOC constants are defined as  $g_{_{l}}=\frac{\partial \omega_c}{\partial x} \sqrt{\frac{\hbar}{m \omega_m}}$ and  $g_{_{q}}=\frac{\partial^2 \omega_c}{\partial x^2} \frac{\hbar}{2m \omega_m}$, respectively with $m$ being the effective mass of the oscillator. 
 The last term describes the interaction of the cavity mode with pump field amplitude 
$(\varepsilon=\sqrt{\frac{2\kappa \mathcal{P}}{\hbar \omega_p}})$, $\mathcal{P}$ being the input power of the pump field. 
 
The description of dynamics and steady state analysis using the linearisation procedure is identical to our previous works \cite{anilkumar,satya}. We consider the respective dissipation terms $\zeta(t)$ and derive Heisenberg equations of motion using Hamiltonian : \begin{equation}
\dot{\mathbf{O}}=\frac{i}{\hbar}[\mathbf{H},\mathbf{O}]+\zeta(t),
\label{s1}
\end{equation} where $\mathbf{O}=(\mathbf{x},\mathbf{p},\mathbf{a})$ and  $\zeta(t)=(0,\xi(t),\sqrt{2\kappa}a_{in})$.
The noise in the field and mechanical mode is described by noise operator $\mathbf{a_{in}}$ with zero mean and $\delta a_{in}(t)$ fluctuations around it and  the Brownian stochastic force described by $\xi(t)$ with zero-mean, having a damping rate $\gamma_m$, respectively. The correlation function at temperature T for these noises are  \small
\begin{subequations} \begin{eqnarray}
 &&\langle \xi(t)\xi(t')\rangle= \frac{\gamma_m}{2\pi\omega_m}\int d\omega e^{-i\omega(t-t')}\omega\left[1+\coth\left(\frac{\hbar\omega}{2k_BT} \right)\right], \\
 && \langle \delta a_{in}(t)\delta a_{in}^\dagger(t')\rangle =(n_a+1)\delta (t-t'),\\
 &&\langle \delta a_{in}^\dagger(t)\delta a_{in}(t')\rangle =n_a\delta (t-t').
\end{eqnarray}\label{a13}\end{subequations} \normalsize where  $k_B$ is the Boltzmann constant and $n_a=\left[ \exp(\frac{\hbar\omega_c}{k_BT})-1\right]^{-1}$ is the equilibrium mean thermal photon number. At optical frequencies $\hbar \omega_c \gg k_BT$ and therefore $n_a\approx 0$.
%\section{Linearisation and steady-state analysis}

We rewrite the Heisenberg operators as complex numbers  with $\langle \mathbf{O} \rangle \equiv O$, representing their respective steady state values
 with the inclusion of fluctuations around their steady state values,
 $i.e$ $O(t)=O_s+\lambda \delta O(t)$. Thus by expanding Eqn.(\ref{s1}) as described above yields,  \small
 \begin{subequations}\begin{eqnarray}
 &&  x_s= \frac{-g_{_{l}}|a_s|^2}{\omega_m+2 g_{_{q}}|a_s|^2} ,
\label{a4}\\
  &&p_s=0,\label{a3}\\
  &&  a_s  =\frac{\varepsilon}{ \kappa+i\left( \Delta+g_{_{l}}x_s +g_{_{q}}   (x^2)_s \right)}.
 \end{eqnarray}\label{a5}\end{subequations}\normalsize
 
Since the fluctuations are assumed to be smaller than the respective steady state values, we linearise the dynamics in fluctuations to an $\mathcal{O}(\lambda)$, giving us the linearised Langevin equations, 
 \begin{equation}\dot{u}(t)=A u(t)+\nu(t),\label{a12}
\end{equation} with column vector of fluctuations in the system being $u(t)^T= \left(\begin{matrix} \delta x(t),\delta p(t),\delta X(t),\delta Y(t) \end{matrix}\right)$ and column vector of noise being 
$\nu(t)^T=\left(\begin{matrix}0,\xi(t),\sqrt{2\kappa}\delta X_{in}(t),\sqrt{2\kappa}\delta Y_{in}(t)\end{matrix}\right)$. Here
 $\delta X \equiv \frac{\delta a+\delta a^\dagger}{\sqrt{2}}$, $\delta Y \equiv \frac{\delta a-\delta a^\dagger}{\sqrt{2}i}$ and their corresponding noises are $\delta X_{in}$ and $\delta Y_{in}$. The matrix $A$ is given by
\small
\begin{eqnarray}A=\left(\begin{matrix} 0&\omega_m &0&0\\ -\tilde{\omega}_m& -\gamma_m& 
-\tilde{G} X_s&-\tilde{G} Y_s\\ \tilde{G} Y_s& 0&-\kappa  & \tilde{\Delta}\\ -\tilde{G} X_s & 0& -\tilde{\Delta}&-\kappa \end{matrix}\right) , \label{matrix_A}
\end{eqnarray}\normalsize
with  $I\equiv |a_s|^2$, $\tilde{\omega}_m \equiv \omega_m+2g_{_{q}} I$, $\tilde{\Delta}\equiv \Delta+g_{_{l}}x_s +g_{_{q}} x_s^2 $, $X_s=\frac{a_s+a_s^*}{\sqrt{2}}$, $Y_s=\frac{a_s-a_s^*}{\sqrt{2}i}$ and
$\tilde{G} \equiv g_{_{l}}  +2 g_{_{q}}x_s$. 
The solutions of Eqn.(\ref{a12}) are stable only if all the eigenvalues of the matrix $A$ have negative real parts.

\subsection{Effect of QOC on the stability of the system}
 The LOC and QOC together in the system provides a non-linear interaction between the cavity light field and mechanical motion. The presence of QOC can modify the dynamics of stability that exists in the system. This can be deduced by applying Routh-Hurwitz criterion \cite{routh} using $A$ given in Eqn.(\ref{matrix_A}) yielding us the following conditions in terms of system parameters:
\small
\begin{subequations}
\begin{eqnarray}
&&s_1 \equiv (\kappa^2+\tilde{\Delta}^2)+2 \kappa \gamma_m+\tilde{\omega}_m \omega_m >0,\\
&& s_2 \equiv (\kappa^2+\tilde{\Delta}^2)\gamma_m+2 \kappa \tilde{\omega}_m \omega_m>0,\\
&&s_3 \equiv (\kappa^2+\tilde{\Delta}^2)\tilde{\omega}_m \omega_m-\tilde{\Delta} \omega_m \tilde{G}^2 (X_s^2+P_s^2)>0,\\
&& (2 \kappa +\gamma_m) s_1 > s_2, \\
&& s_1 s_2 (2 \kappa +\gamma_m)> s_2^2+(2 \kappa +\gamma_m)^2 s_3.
\end{eqnarray}\label{a15}\end{subequations}\normalsize  

With the inclusion of QOC in the regular OM system, the effective spring constant of the mechanical oscillator $\mathcal{K}=m\omega_m^2$ modifies to $\mathcal{\tilde{K}}=m\omega_m\tilde{\omega}_m$ , as estimated from Eqn.(\ref{s1})\footnote{Note that Eqn.(\ref{s1}) is expressed in dimensionless quantities. $\mathcal{K}$ is deduced by converting dimensionless $\mathbf{x,p}$ to $\mathbf{\hat{x},\hat{p}}$ with dimensions, using $\mathbf{x}=\mathbf{\hat{x}}\sqrt{m\omega_m/\hbar}$ and $\mathbf{p}=\mathbf{\hat{p}}\sqrt{1/m\omega_m\hbar}$.} and the effective OM coupling from $g_l$ to $\tilde{G}=g_l+g_q x_s$.  The positive (negative) QOC stiffens (softens) the oscillator in comparison to the case of LOC alone. The modified spring constant brings a change in the restoring force, and hence to balance it, the radiation pressure given by $F_{rad}= (\hbar \omega_c/L)\langle a^\dagger a\rangle\propto\frac{\mathcal{P}}{\tilde{\Delta}^2+\kappa^2}$ has to readjust thereafter. In cases where it is not possible to achieve this, the system becomes unstable and results in not satisfying the Routh-Hurwitz criteria given in Eqn.(\ref{a15}). The stability region of such a system with positive QOC extends while it gets shrunk with negative QOC as studied extensively in \cite{satya,anilkumar}. However, the negative QOC favours a strong OM interaction which can be quantified by studying NMS. The following section provide us with the necessary theoretical analysis of NMS.

\section{Fluctuations and analysis of normal mode splitting}
Using the definition of Fourier transform, $\mathcal{F}(\omega)=\frac{1}{2\pi}\int_{-\infty}^\infty\mathcal{F}(t)e^{-i\omega t}dt$ and $[\mathcal{F}^\dagger(\omega)]^\dagger=\mathcal{F}(-\omega)$  in the Eqn.(\ref{a12}), the set of coupled differential equations form a simple system of linear equations in frequency.  Therefore after solving the matrix equation Eqn.(\ref{a12}) in frequency domain, we get 
\small
\begin{equation}
\delta x (\omega)=\frac{1}{D(\omega)}\lbrace X_a(\omega) a_{in}(\omega)+ X_{a^\dagger}(\omega) a_{in}^\dagger (\omega) -X_{\xi}(\omega)\xi(\omega) \rbrace ,\label{b0}
\end{equation}\normalsize where \small
\begin{subequations}
\begin{eqnarray}
&&D(\omega)= \left((\kappa -i \omega)^2+\tilde{\Delta}^2\right)\left( \omega^2+i \gamma_m \omega- \omega_m \tilde{\omega}_m)\right) \label{c1}\nonumber\\&&+2\tilde{G}^2I \tilde{\Delta}\omega_m ,\\
&&X_a(\omega)=\sqrt{2 \kappa} \omega_m \tilde{G}a_s^* (\kappa - i\omega - 
   i \tilde{\Delta}),\\
&&X_{a^\dagger}(\omega)= (X_a(-\omega))^\dagger, \\
&&X_\xi(\omega)=\omega_m ((\kappa - i \omega)^2 + \tilde{\Delta}^2).
\end{eqnarray}\label{b2}
\end{subequations}\normalsize 
 
The spectrum of fluctuations in position of the movable
mirror is defined by\small
\begin{equation}
S_{xx}(\omega)=\frac{1}{4\pi}\int e^{-i(\omega+\Omega) t} \langle \delta x(\omega)\delta x(\Omega) +\delta x(\Omega)\delta x(\omega)\rangle d\Omega.\label{a16}
\end{equation}
\normalsize
The required correlations of noise operators given in Eqn.(\ref{a13}) are used in their frequency domain as: 

\begin{subequations}
\begin{eqnarray}
&&\langle \delta a_{in}(\omega)\delta a_{in}^\dagger(\omega')\rangle =2\pi\delta(\omega+\omega'),\\
&&\langle \xi(\omega)\xi(\omega')\rangle = 2\pi \frac{\omega\gamma_m}{\omega_m}\left[\coth\left(\frac{\hbar \omega}{2k_BT}\right)+1\right]\delta(\omega+\omega').\nonumber\\
\end{eqnarray}\label{b4}\end{subequations}\normalsize
Substituting Eqn.(\ref{b0}) and Eqn.(\ref{b4}) in Eqn.(\ref{a16}), we obtain the spectrum of fluctuations in position of the movable mirror, 

\small
\begin{eqnarray}
&&S_{xx}(\omega)=\frac{\omega_m}{|D(\omega^2)|}\left[2\tilde{G}^2I\kappa\omega_m(\kappa^2+\omega^2+\tilde{\Delta}^2)\right.\nonumber\\&&\left.+\omega\gamma_m\coth\left(\frac{\hbar\omega}{k_BT}\right)\left(\left(\kappa^2+\omega^2+\tilde{\Delta}^2\right)^2 -4\omega^2\tilde{\Delta}\right)\right]
\label{sq}
\end{eqnarray}
\normalsize
The terms proportional to $\tilde{G}$ and  $\xi(\omega)$ in Eqn.(\ref{b0}) describes the effect of the radiation pressure and  thermal noise on mirror's motion, respectively.\hspace*{+0.15cm}In case of no OM coupling, we have $\delta x(\omega)= \omega_m\xi(\omega)/\left(\omega_m^2-\omega^2-i\gamma_m\omega\right)$ where the movable mirror makes Brownian motion, whose susceptibility has a Lorentzian shape centred at frequency $\omega_m$ with width $\gamma_m$. But now due to LOC and QOC, both thermal noise and radiation pressure decide the mechanical susceptibility that can reveal interesting features as discussed in the following section.

\section{Numerical results}
In all our numerical calculations we scale QOC values with LOC value shown as $g_{_{q}}/g_{_{l}}$. The parameters chosen in our calculations are similar to those used in \cite{groblacher2009}. The cavity  with a linewidth of $\kappa/2\pi= 215$ kHz is driven by a laser of wavelength 1064 nm. The OM system works in the resolved side-band regime i.e. $\omega_m>\kappa$ such that the mechanical oscillator  has a frequency $ \omega_m/2\pi=947$ kHz with a  damping rate $\gamma_m/2\pi=141.34$ Hz. The free mass of the oscillator is $m=145$ ng and it interacts with the optical field through  LOC, $g_{_{l}}/2\pi = 3.95$ Hz. The system is assumed to be in contact with a  thermal bath at 300 mK. 

 Usually, the strong interaction between  optical light circulating inside the cavity and movable mirror through LOC exhibits NMS \cite{dobrindt,groblacher2009} effect in the resolved side band regime. NMS arises due to photon-phonon interactions, wherein the energy exchange among them occurs in time-scales faster than their individual decoherence rates. To investigate the effect of QOC on NMS, in its magnitude and parity, we have calculated the mechanical oscillator's position spectrum, $S_{xx}(\omega)$. 
%\begin{widetext}
\begin{figure*}[t]
\framebox{\includegraphics[scale=0.34]{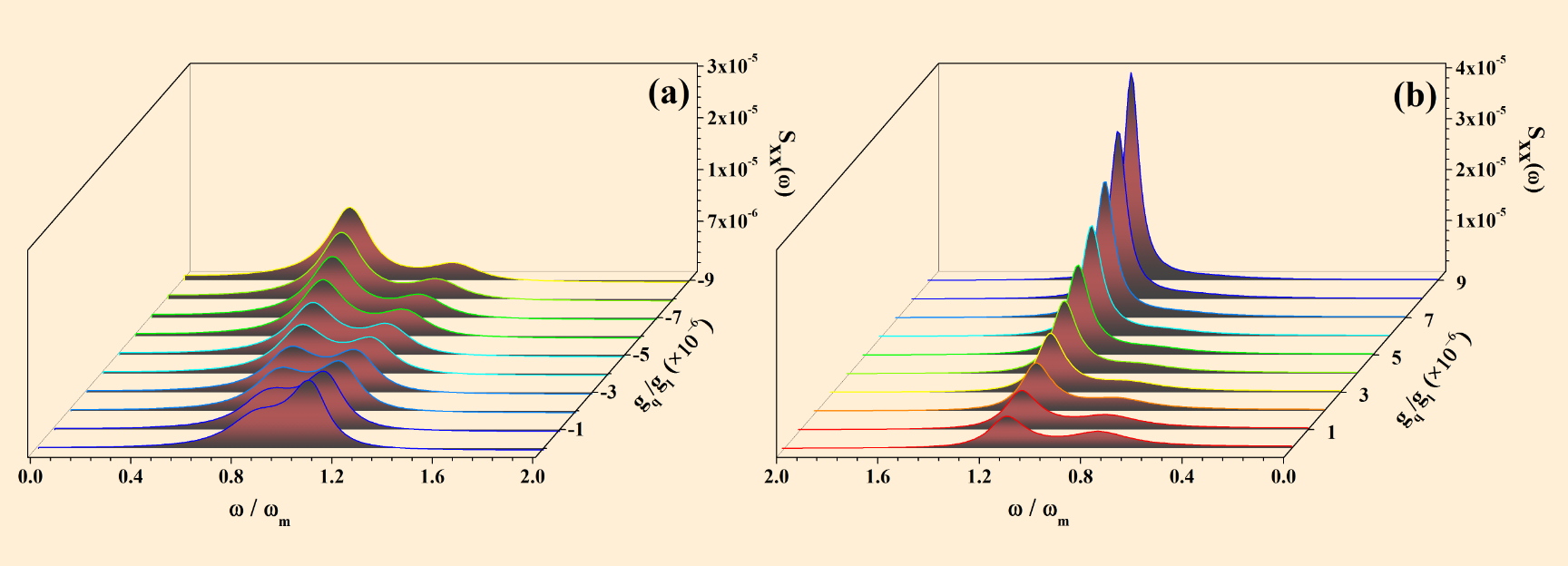}}
\caption{(a) and (b) Shows the mechanical oscillator's position spectrum $S_{xx}(\omega)$  plotted as a function of normalized frequency ($\omega/\omega_m$) and QOC ($g_q/g_l)$. The system parameters are  $\Delta=\omega_m$ ,$T = 300$ mK  and input powers are 6.9 mW for (a) and 10.7 mW for (b), respectively. }
\end{figure*}
%\end{widetext}

Figure. 2(a) and (b) shows the calculated $S_{xx}(\omega)$  plotted for an input powers of 6.9 mW and 10.7 mW as a function of normalised frequency $(\omega/\omega_m )$ for both negative and positive QOC ($g_q /g_l$), respectively . The QOC value is varied from $-9\times10^{-6}$ to $9\times10^{-6}$. It can be seen that in the absence of QOC, i.e. $g_q/g_l=0$ and at moderate powers (6.9 mW), the spectrum exhibit a partially resolved peaks. The observed NMS is asymmetric in nature and a better spectral resolution is evident for QOC values beyond $-4 \times 10^{-6}$.  On the contrary in Fig. 2(b), we evaluate the spectrum at 10.7 mW exhibits a well-resolved NMS peaks. However, it could be seen that on increasing the positive value of QOC further, one of the peaks grow relatively stronger than the other, thus eventually leading to a single peak structure. A well-resolved double peak structure would imply a stronger NMS. Hence, the presence of a negative QOC enhances the NMS in a conventional OM system, while the positive QOC inhibits it.

Further analysis of position and width of the NMS peaks was analysed by studying $D(\omega)$ given in Eqn.(\ref{c1}). The position and width of the NMS peaks can be estimated from the zeros of $D(\omega)$. While the real part gives the information of the position, the corresponding imaginary part gives its width. Being an expression in $\omega^4$, the solutions of $D(\omega)$ are complex conjugates of two solutions expressed as $\omega_{+} \pm i \Gamma_+$ and $\omega_{-} \pm i \Gamma_-$, such that $\omega_+>\omega_m>\omega_-$. These values are calculated numerically, normalised with respect to $\omega_m$ and are plotted as a function of QOC in Fig.3. Figure 3(a)/(b) shows position (width) of the peaks that occur at frequencies centred around $\omega_- (\Gamma_-)$ and around $\omega_+ (\Gamma_+)$. The solid ($\omega_-$) and dashed blue ($\omega_+$) curves correspond to the case of negative QOC whereas the thick solid ($\omega_-$) and dashed ($\omega_+$) red curve corresponds to positive QOC, evaluated at 6.9 mW. In order to also depict the degradation of NMS in Fig. 2(b), we evaluated the roots at 10.7 mW  shown in red thin dashed ($\omega_+$) and solid ($\omega_-$) curves with markers. At 10.7 mW, the stability region for negative QOC does not exist and hence we only plotted for the positive QOC. 
%\begin{widetext}
\begin{figure*}[t]
\hspace*{-0.5cm}\includegraphics[scale=.4]{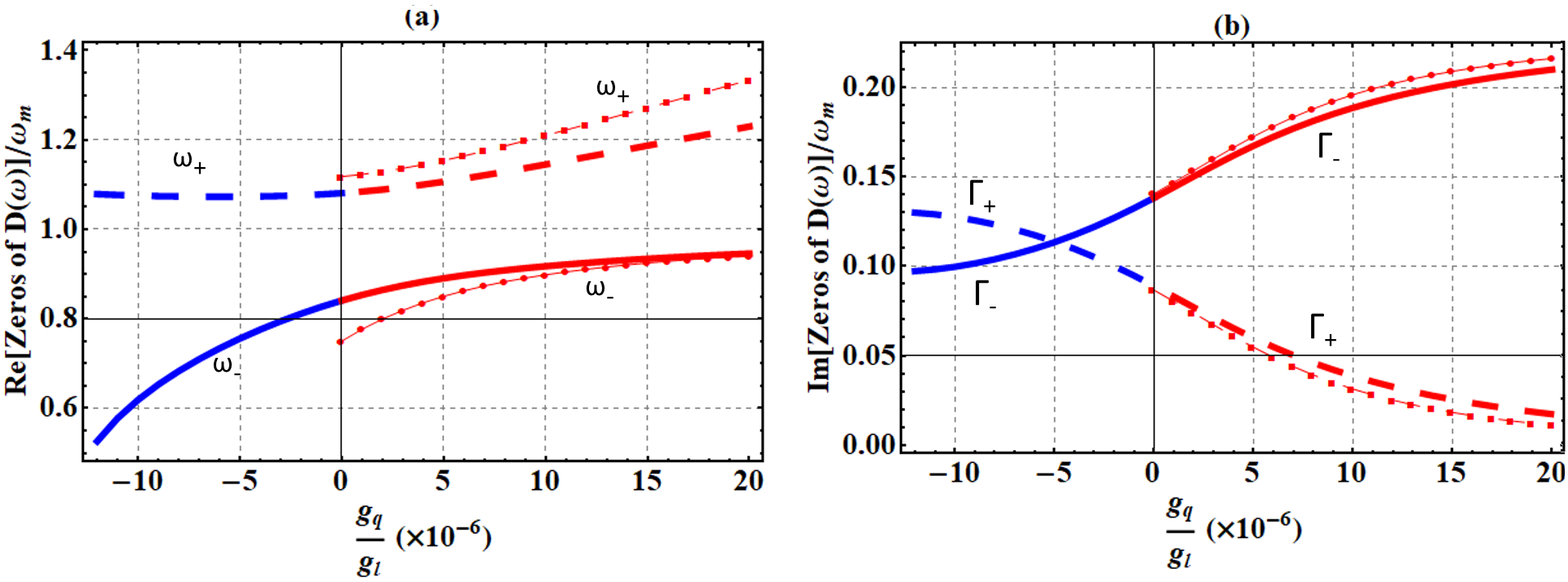}
\caption{Shows (a)position and (b)width of NMS peaks calculated using the real and imaginary part of zeros of $D(\omega)$ given in Eqn.(\ref{c1}) as a function of QOC ($g_q/g_l$). The blue color corresponds to negative QOC value and red corresponds to the positive QOC value. The thick solid line and dashed line (both blue and red) are plotted for 6.9 mW, where as the thin (solid and dashed) curves with markers are plotted at 10.7 mW. There lies no stable region for negative QOC at 10.7 mW and therefore blue curves are not shown. }
\end{figure*}
%\end{widetext}

In Fig.3, we chose the values of QOC ($g_q/g_l$) ranging from $-12\times 10^{-6}$ to $20\times 10^{-6}$.  For an increasing negative QOC, the separation between the peaks as shown with blue curve in Fig.3(a) gets larger as compared to the variation seen in their widths in Fig. 3(b). This leads to a stronger NMS. In the case of positive QOC, the spacing between the relative peak positions increase with QOC as shown by red curves in Fig. 3(a). But one of the peak widths ($\Gamma_-$) in Fig.3(b) increases at a faster rate leading to the disappearance of the double-peak structure. The dominant peak in either of the cases depend on the parity of QOC. The peak amplitude around $\omega_-$ is larger in case of negative QOC and vice versa for positive QOC as shown in Fig.2.  

Enhanced NMS suggests a stronger OM coupling which can be understood by writing an explicit analytic expression for the peak positions using the approximation $\kappa \gg \gamma_m$ and $\tilde{\Delta} \gg \gamma_m$:
\small \begin{eqnarray}
&\omega^2_\pm=\frac{1}{2}\left(\omega_a^2+\omega_b^2\right)\pm\frac{1}{2}\sqrt{\left(\omega_a^2-\omega_b^2\right)^2+\left(8\omega_mI\tilde{G}^2\tilde{\Delta}\right)}\label{A1}
\end{eqnarray}\normalsize 
with  $\omega_a^2 \equiv \kappa^2+\tilde{\Delta}^2$ and $\omega_b^2\equiv \omega_m\tilde{\omega}_m$.  In the current scenario of resolved side-band regime we have, $\tilde{\Delta} \approx \omega_m > \kappa$ and $\tilde{\Delta}^2\approx \omega_m\tilde{\omega}_m$. 
\begin{figure}[h!]
\hspace*{-0.5cm}\includegraphics[scale=0.5]{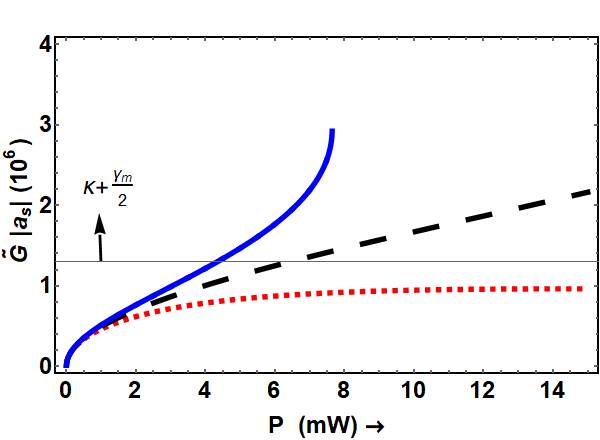}
\caption{The plot shows the position dependent effective OM coupling varied as function of input power for various QOC ($g_q/g_l$) values as   $-12\times 10^{-6}$ (blue solid), 0 (Black dashed) and $20\times10^{-6}$ (Red dotted), respectively. The plot for the blue curve was shown only till 7.6 mW, since the system for negative QOC tends to become unstable at higher powers. The black thin solid line (labelled as $\kappa+\gamma_m/2$) is the sum of  widths of the  peaks in NMS spectra. Regions with effective coupling greater than the black thin solid line shows NMS and vice versa.} 
\end{figure}

The peak separation from the above expression can be estimated to be $\omega_+-\omega_- \propto \tilde{G}|a_s|$.  Hence, a stronger many photon OM coupling will lead to a greater peak separation and thereby enhancing NMS, consistent as shown in the special case of LOC alone in \cite{groblacher2009}. This implies that the energy exchange between the optical and mechanical modes is greater than the decoherence rate. The decoherence in the system manifests  as the width of the peaks which can be estimated by the imaginary part of the solutions of $D(\omega)$ as shown in Appendix section. To illustrate this, we plotted the value of $\tilde{G}|a_s|$ as a function of power in Fig.4  relative to the sum of widths i.e. $\kappa+\gamma_m/2$ (shown in thin black solid curve), with QOC values as 0 (black dashed curve), negative ($g_q/g_l=-12\times 10^{-6}$ with blue solid curve) and positive QOC ( $g_q/g_l=20\times 10^{-6}$ with red dotted curve). NMS is evident for the regions where the values of $\tilde{G}a_s>\kappa+\gamma_m/2$. We can see that the value of $\tilde{G}|a_s|$ at negative QOC is almost thrice the value of its corresponding positive QOC at 7.6 mW, beyond which the system is unstable for negative QOC. Hence by having a handle on the magnitude and parity of QOC could relatively increase or decrease the contribution of radiation pressure to the mechanical susceptibility over thermal noise as shown in Eqn.(\ref{sq}). This would lead to a controlled OM interaction manifested through NMS. 

It is remarkable that a meagre amount of QOC which is 1 in a million parts of LOC could manifest such features. Similar features were observed previously when theoretical studies were undertaken in hybrid-OM systems \cite{huangnms,kumar,naderi} and in this sense, we mean that a system with QOC can mimic a hybrid-OM system. In \cite{naderi}, an OPA and a Kerr medium were placed inside a cavity OM system in order to control the NMS. While NMS got enhanced under a stronger OPA effect, the Kerr effect deteriorated NMS. These features are very close to the current scenario when the QOC's parity is changed from negative to positive. 

\begin{figure}
\hspace*{-0.5cm}\includegraphics[scale=0.7]{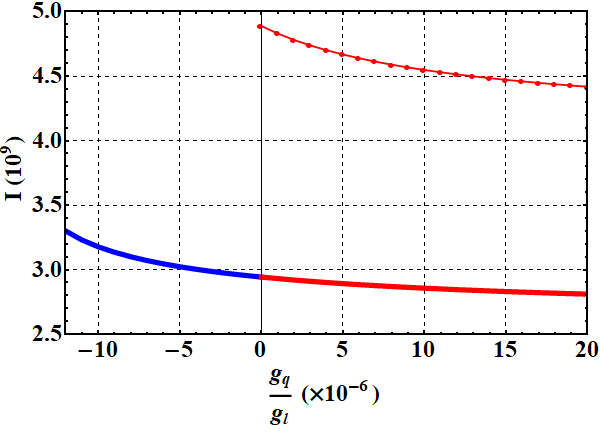}
\caption{ The plot shows photons $ \langle a^{\dagger}a\rangle = I$ circulating inside the cavity varied as function of QOC ($g_q/g_l$) at 6.9 mW (solid red and blue curve) and 10.7 mW (red thin curve with markers). }
\end{figure}

The physical explanation leading to such a phenomenon is given as follows. The presence of QOC modifies the spring constant from  $\mathcal{K}$ to $\tilde{\mathcal{K}}$. The negative QOC softens the spring while the positive QOC stiffens it. The softened spring constant aids in accommodating more number of intracavity photons ($I$) as shown in Fig. 5 with blue color as compared to with $g_q/g_l=0$. As a result, this also leads to an increase in the effective many photon OM coupling rate ($\tilde{G}|a_s|$) as shown by the blue curve in Fig.4. Therefore the system acts as a gain medium in presence of negative QOC and mimics a hybrid-OM system containing an OPA \cite{huangnms} that creates the excess photons. In contrast, the stiffening of spring decreases the number of intracavity photons as shown in Fig.5 evaluated at 6.9 mW (red solid curve) and 10.7 mW (thin red curve with markers). This decreases the effective many photon OM coupling as shown with a red dotted curve in Fig.4. This leads to an inhibition of NMS and can be identified by a hybrid-OM system containing a Kerr medium as proposed in \cite{kumar}. Both \cite{kumar,naderi} have concluded that the presence of Kerr medium invokes the photon-blockade mechanism by which NMS can be controlled. Hence, our proposal could possibly open avenues where the photon blockade mechanism itself could be controlled using the strength and parity of QOC.

\section{Conclusion}   
We have successfully demonstrated the effect of QOC on normal-mode splitting over a conventional OM system. Despite the QOC being too small, we could show that the features it can manifest in an OM system were quite significant. The presence of such a non-linearity with a freedom in controlling its parity could increase or decrease the intracavity photons and thereby the effective OM coupling rate. This could lead to either enhancement or deterioration on normal-mode splitting as found in the case of hybrid-OM systems \cite{huangnms,kumar,naderi}. Hence, systems with LOC and QOC could prove to be a promising alternative platforms to hybrid-OM systems wherein one could see quantum features at possibly higher bath temperatures without introducing any extra degrees-of-freedom. The experimental realization of both LOC and QOC were observed in various systems such as  a nano and micro mechanical resonators \cite{schwab,hertzberg},  fiber based optical cavity with silicon nitride membrane \cite{flowers}, micro disk resonators \cite{doolin} and  levitated nanoparticles \cite{fonseca}, making our scheme feasible.

\section*{Acknowledgement(s)}
M.A.K would like to thank the computing facilities provided by NIT, Andhra Pradesh. U.S.S. is supported by GUPRS and GUIPRS. The authors would also like to thank Kavya Hemantha Rao for her contributions with the presentation of the figures.

\section*{Disclosure statement}
No potential conflict of interest was reported by the authors.

\section*{Appendix: Estimating the widths of $S_{xx}(\omega)$}
The peak positions and widths in $S_{xx}(\omega)$ can be estimated from the real and imaginary part of the zeros of $D(\omega)$, respectively. Since $D(\omega)$ is a quartic equation, we can analytically solve using various methods \cite{abramowitz1965}, the expression looks cumbersome and does not really help the reader. While the real part was estimated, in the current scenario by  Eqn.(\ref{A1}), the complex part of the solution is not so straightforward. However it is known from our numerical solutions that the zeros of $D(\omega)$ are of the form $-\omega_{\pm}\pm\Gamma_{\pm}$, where $\omega_{\pm}$ is the real part ( $ \omega_\pm$ in Eqn.(\ref{A1})) and $\Gamma_{\pm}$ is the imaginary part of the solutions. Hence we can write
\small
\begin{eqnarray*}
D(\omega)=[\omega-(-\omega_{+}+i\Gamma_{+})][\omega -(-\omega_{+}-i\Gamma_{+})][\omega-(-\omega_{-}+i\Gamma_{-})][\omega-(-\omega_{-}-i\Gamma_{-})]
\end{eqnarray*}
 \normalsize
By straightforward algebraic expansion, one find that the coefficient of the third order term, $\omega^3$ is the sum of imaginary part of the solutions i.e. $\Gamma_{+}+\Gamma_{-}=2\kappa+\gamma_m$. Since the imaginary parts correspond to the widths of the peaks in position spectra, $2\kappa+\gamma_m$ is the sum of widths of the Lorentzians in $S_{xx}(\omega)$. A well resolved double-peak structure implies that the spacing between the peaks is greater than the sum of (half) widths of the peaks, i.e. $\omega_{+}-\omega_{-}>(\Gamma_{+}+\Gamma_{-})/2$.
\bibliographystyle{tfp}
\bibliography{ref}

\begin{thebibliography}{10}
\providecommand{\url}[1]{\normalfont{#1}}
\providecommand{\urlprefix}{}

\bibitem{aspelmeyer}
Aspelmeyer, M.; Kippenberg, T.J.; Marquardt, F. Cavity optomechanics,
  \emph{Rev. Mod. Phys.}  \textbf{2014}, \emph{86}~(4), 1391.

\bibitem{meystre}
Meystre, P. A short walk through quantum optomechanics, \emph{Ann. Phys.
  (Berlin)}  \textbf{2013}, \emph{525}~(3), 215--233.

\bibitem{arcizet}
Arcizet, O.; Cohadon, P.F.; Briant, T.; Pinard, M.; Heidmann, A.
  Radiation-pressure cooling and optomechanical instability of a micromirror,
  \emph{Nature (London)}  \textbf{2006}, \emph{444}~(7115), 71--74.

\bibitem{dobrindt}
Dobrindt, J.M.; Wilson-Rae, I.; Kippenberg, T.J. Parametric Normal-Mode
  Splitting in Cavity Optomechanics, \emph{Phys. Rev. Lett.}  \textbf{2008},
  \emph{101}, 263602.
  \urlprefix\url{http://link.aps.org/doi/10.1103/PhysRevLett.101.263602}.

\bibitem{groblacher2009}
Gr{\"o}blacher, S.; Hammerer, K.; Vanner, M.R.; Aspelmeyer, M. Observation of
  strong coupling between a micromechanical resonator and an optical cavity
  field, \emph{Nature (London)}  \textbf{2009}, \emph{460}~(7256), 724.

\bibitem{rossi}
Rossi, M.; Kralj, N.; Zippilli, S.; Natali, R.; Borrielli, A.; Pandraud, G.;
  Serra, E.; Di~Giuseppe, G.; Vitali, D. Normal-Mode Splitting in a Weakly
  Coupled Optomechanical System, \emph{Phys. Rev. Lett.}  \textbf{2018},
  \emph{120}, 073601.
  \urlprefix\url{https://link.aps.org/doi/10.1103/PhysRevLett.120.073601}.

\bibitem{eberly}
Sanchez-Mondragon, J.J.; Narozhny, N.B.; Eberly, J.H. Theory of
  Spontaneous-Emission Line Shape in an Ideal Cavity, \emph{Phys. Rev. Lett.}
  \textbf{1983}, \emph{51}, 550--553.
  \urlprefix\url{http://link.aps.org/doi/10.1103/PhysRevLett.51.550}.

\bibitem{thompson}
Thompson, R.J.; Rempe, G.; Kimble, H.J. Observation of normal-mode splitting
  for an atom in an optical cavity, \emph{Phys. Rev. Lett.}  \textbf{1992},
  \emph{68}, 1132--1135.
  \urlprefix\url{http://link.aps.org/doi/10.1103/PhysRevLett.68.1132}.

\bibitem{agarwal}
Agarwal, G.S. Vacuum-Field Rabi Splittings in Microwave Absorption by Rydberg
  Atoms in a Cavity, \emph{Phys. Rev. Lett.}  \textbf{1984}, \emph{53},
  1732--1734.
  \urlprefix\url{http://link.aps.org/doi/10.1103/PhysRevLett.53.1732}.

\bibitem{hybrid}
Rogers, B.; Gullo, N.L.; De~Chiara, G.; Palma, G.M.; Paternostro, M. Hybrid
  optomechanics for quantum technologies, \emph{Quantum Meas. Quantum Metrol.}
  \textbf{2014}, \emph{2}~(1).

\bibitem{genes2008}
Genes, C.; Vitali, D.; Tombesi, P. Emergence of atom-light-mirror entanglement
  inside an optical cavity, \emph{Physical Review A}  \textbf{2008},
  \emph{77}~(5), 050307.

\bibitem{genes2011}
Genes, C.; Ritsch, H.; Drewsen, M.; Dantan, A. Atom-membrane cooling and
  entanglement using cavity electromagnetically induced transparency,
  \emph{Physical Review A}  \textbf{2011}, \emph{84}~(5), 051801.

\bibitem{wang2014}
Wang, H.; Gu, X.; Liu, Y.x.; Miranowicz, A.; Nori, F. Optomechanical analog of
  two-color electromagnetically induced transparency: Photon transmission
  through an optomechanical device with a two-level system, \emph{Physical
  Review A}  \textbf{2014}, \emph{90}~(2), 023817.

\bibitem{chen}
Chen, A.; Nie, W.; Li, L.; Zeng, W.; Liao, Q.; Xiao, X. Steady-state
  entanglement in levitated optomechanical systems coupled to a higher order
  excited atomic ensemble, \emph{Optics Communications}  \textbf{2017},
  \emph{403}, 97--102.

\bibitem{huangnms}
Huang, S.; Agarwal, G.S. Normal-mode splitting in a coupled system of a
  nanomechanical oscillator and a parametric amplifier cavity, \emph{Phys. Rev.
  A}  \textbf{2009}, \emph{80}, 033807.
  \urlprefix\url{http://link.aps.org/doi/10.1103/PhysRevA.80.033807}.

\bibitem{kumar}
Kumar, T.; Bhattacherjee, A.B.; ManMohan. Dynamics of a movable micromirror in
  a nonlinear optical cavity, \emph{Phys. Rev. A}  \textbf{2010}, \emph{81},
  013835. \urlprefix\url{https://link.aps.org/doi/10.1103/PhysRevA.81.013835}.

\bibitem{naderi}
Shahidani, S.; Naderi, M.; Soltanolkotabi, M. Normal-mode splitting and
  output-field squeezing in a Kerr-down conversion optomechanical system,
  \emph{Journal of Modern Optics}  \textbf{2015}, \emph{62}~(2), 114--124.

\bibitem{thompson2008}
Thompson, J.; Zwickl, B.; Jayich, A.; Marquardt, F.; Girvin, S.; Harris, J.
  Strong dispersive coupling of a high-finesse cavity to a micromechanical
  membrane, \emph{Nature (London)}  \textbf{2008}, \emph{452}~(7183), 72.

\bibitem{sankey}
Sankey, J.C.; Yang, C.; Zwickl, B.M.; Jayich, A.M.; Harris, J.G. Strong and
  tunable nonlinear optomechanical coupling in a low-loss system, \emph{Nat.
  Phys.}  \textbf{2010}, \emph{6}~(9), 707--712.

\bibitem{purdy}
Purdy, T.P.; Brooks, D.W.C.; Botter, T.; Brahms, N.; Ma, Z.Y.; Stamper-Kurn,
  D.M. Tunable Cavity Optomechanics with Ultracold Atoms, \emph{Phys. Rev.
  Lett.}  \textbf{2010}, \emph{105}, 133602.
  \urlprefix\url{http://link.aps.org/doi/10.1103/PhysRevLett.105.133602}.

\bibitem{kaviani}
Kaviani, H.; Healey, C.; Wu, M.; Ghobadi, R.; Hryciw, A.; Barclay, P.E.
  Nonlinear optomechanical paddle nanocavities, \emph{Optica}  \textbf{2015},
  \emph{2}~(3), 271--274.
  \urlprefix\url{http://www.osapublishing.org/optica/abstract.cfm?URI=optica-2-3-271}.

\bibitem{brawley}
Brawley, G.; Vanner, M.; Larsen, P.E.; Schmid, S.; Boisen, A.; Bowen, W.
  Nonlinear optomechanical measurement of mechanical motion, \emph{Nat.
  Commun.}  \textbf{2016}, \emph{7}, 10988.

\bibitem{paraso}
Para\"{\i}so, T.K.; Kalaee, M.; Zang, L.; Pfeifer, H.; Marquardt, F.; Painter,
  O. Position-Squared Coupling in a Tunable Photonic Crystal Optomechanical
  Cavity, \emph{Phys. Rev. X}  \textbf{2015}, \emph{5}, 041024.
  \urlprefix\url{http://link.aps.org/doi/10.1103/PhysRevX.5.041024}.

\bibitem{girvin}
Nunnenkamp, A.; B\o{}rkje, K.; Harris, J.G.E.; Girvin, S.M. Cooling and
  squeezing via quadratic optomechanical coupling, \emph{Phys. Rev. A}
  \textbf{2010}, \emph{82}, 021806.
  \urlprefix\url{http://link.aps.org/doi/10.1103/PhysRevA.82.021806}.

\bibitem{tan2013}
Tan, H.; Bariani, F.; Li, G.; Meystre, P. Generation of macroscopic quantum
  superpositions of optomechanical oscillators by dissipation, \emph{Physical
  Review A}  \textbf{2013}, \emph{88}~(2), 023817.

\bibitem{liao2013}
Liao, J.Q.; Nori, F.; et~al. Photon blockade in quadratically coupled
  optomechanical systems, \emph{Physical Review A}  \textbf{2013},
  \emph{88}~(2), 023853.

\bibitem{xu2018}
Xu, X.W.; Shi, H.Q.; Chen, A.X.; Liu, Y.x. Cross-correlation between photons
  and phonons in quadratically coupled optomechanical systems, \emph{Phys. Rev.
  A}  \textbf{2018}, \emph{98}, 013821.
  \urlprefix\url{https://link.aps.org/doi/10.1103/PhysRevA.98.013821}.

\bibitem{anilkumar}
U, S.S.; Anil~Kumar, M. Effects of linear and quadratic dispersive couplings on
  optical squeezing in an optomechanical system, \emph{Phys. Rev. A}
  \textbf{2015}, \emph{92}, 033824.
  \urlprefix\url{https://link.aps.org/doi/10.1103/PhysRevA.92.033824}.

\bibitem{satya}
Satya~Sainadh, U.; Anil~Kumar, M. Squeezing of the mechanical motion and
  beating 3 dB limit using dispersive optomechanical interactions, \emph{J.
  Mod. Opt.}  \textbf{2017}, \emph{64}~(12), 1121--1128.

\bibitem{routh}
DeJesus, E.X.; Kaufman, C. Routh-Hurwitz criterion in the examination of
  eigenvalues of a system of nonlinear ordinary differential equations,
  \emph{Phys. Rev. A}  \textbf{1987}, \emph{35}, 5288--5290.

\bibitem{schwab}
Rocheleau, T.; Ndukum, T.; Macklin, C.; Hertzberg, J.; Clerk, A.; Schwab, K.
  Preparation and detection of a mechanical resonator near the ground state of
  motion, \emph{Nature (London)}  \textbf{2010}, \emph{463}~(7277), 72--75.

\bibitem{hertzberg}
Hertzberg, J.; Rocheleau, T.; Ndukum, T.; Savva, M.; Clerk, A.; Schwab, K.
  Back-action-evading measurements of nanomechanical motion, \emph{Nature
  Physics}  \textbf{2010}, \emph{6}~(3), 213.

\bibitem{flowers}
Flowers-Jacobs, N.; Hoch, S.; Sankey, J.; Kashkanova, A.; Jayich, A.; Deutsch,
  C.; Reichel, J.; Harris, J. Fiber-cavity-based optomechanical device,
  \emph{Appl. Phys. Lett.}  \textbf{2012}, \emph{101}~(22), 221109.

\bibitem{doolin}
Doolin, C.; Hauer, B.D.; Kim, P.H.; MacDonald, A.J.R.; Ramp, H.; Davis, J.P.
  Nonlinear optomechanics in the stationary regime, \emph{Phys. Rev. A}
  \textbf{2014}, \emph{89}, 053838.
  \urlprefix\url{http://link.aps.org/doi/10.1103/PhysRevA.89.053838}.

\bibitem{fonseca}
Fonseca, P.Z.G.; Aranas, E.B.; Millen, J.; Monteiro, T.S.; Barker, P.F.
  Nonlinear Dynamics and Strong Cavity Cooling of Levitated Nanoparticles,
  \emph{Phys. Rev. Lett.}  \textbf{2016}, \emph{117}, 173602.
  \urlprefix\url{https://link.aps.org/doi/10.1103/PhysRevLett.117.173602}.

\bibitem{abramowitz1965}
Abramowitz, M.; Stegun, I.A. \emph{Handbook of mathematical functions: with
  formulas, graphs, and mathematical tables}; Courier Corporation, 1965;
  Vol.~55.

\end{thebibliography}
\end{document}